\documentclass{revtex4}
\usepackage{amssymb}
\usepackage{amsmath}

\input{tcilatex}

\begin{document}

\title[ ]{Quantum singularities in a model of $f(R)$ gravity}
\author{O. Gurtug}
\email{ozay.gurtug@emu.edu.tr}
\author{T. Tahamtan}
\email{tayabeh.tahamtan@emu.edu.tr}
\affiliation{Department of Physics, Eastern Mediterranean University, G. Magusa, north
Cyprus, Mersin 10, Turkey. }

\begin{abstract}
The formation of a naked singularity in a model of  $f(R)$ gravity having as
source a linear electromagnetic field is considered in view of quantum
mechanics. Quantum test fields obeying the Klein$-$Gordon, Dirac and Maxwell
equations are used to probe the classical timelike naked singularity
developed at $r=0$. We prove that the spatial derivative operator of the
fields fails to be essentially self-adjoint. As a result, the classical
timelike naked singularity remains quantum mechanically singular when it is
probed with quantum fields having different spin structures.
\end{abstract}

\maketitle

\section{Introduction}

In the last decade, there have been extensive studies in Extended Theories
of Gravity (ETG) such as the Lovelock and $f(R)$ gravity theories. The main
motivation to study the ETG is to understand the accelerated expansion of
the universe and the issue of dark matter/energy (see \cite{1} and
references therein for a general review). One of the most attractive
branches of the ETG is the $f(R)$ gravity theory in which the standard
Einstein's gravity is extended with an arbitrary function of the Ricci
scalar $R$ instead of the linear one \cite{1}. In this model, the Ricci
scalar $R$ in the Einstein$-$Hilbert action is replaced with $f(R)=R+\alpha
g(R),$ where $g(R)$ is an arbitrary function of $R$ so that in the limit $%
\alpha =0,$ one recovers the Einstein limit. Although the majority of
researchers prefer to use this ansatz, in general, finding an exact analytic
solution to the field equations is not an easy task. As far as analytic
exact solutions are concerned, static, spherically symmetric models in $f(R)$
gravity have been shown to serve for this purpose \cite{2,3,4,5,6}. In this
context of static, spherically symmetric solutions of $f(R)$ gravity, the
solutions admitting black holes have attracted much attention.

In the context of static, spherically symmetric $f(R)$ gravity, it has
recently been shown that \cite{7}, an exact analytic solution is also
possible if one assumes $f(R)$ to have the form of $f(R)=\xi \left(
R+R_{1}\right) +2\alpha \sqrt{R+R_{0}},$ in which $\xi ,\alpha ,R_{0}$ and $%
R_{1}$ are constants, a priority to secure the Einstein limit by setting the
constants $R_{0}=R_{1}=\alpha =0$ and $\xi =1.$ In this model of $f(R)$
gravity, exact solutions with external electromagnetic sources (both linear
and nonlinear) are found. It was shown that the solution with a linear
electromagnetic field does not admit a black hole while the solution with a
nonlinear electromagnetic source admits a black hole solution. The physical
properties of the latter solution are investigated by calculating
thermodynamic quantities and it was shown to satisfy the first law of
thermodynamics. The solution having as a source a linear electromagnetic
field resulted with a naked curvature singularity at $r=0,$ which is a
typical central singularity peculiar to spherically symmetric systems. The
solution given in \cite{7}, is a kind of extension of a global monopole
solution \cite{8} which represents a solution of the Einstein's equations
with spherical symmetry with matter that extends to infinity. It can also be
interpreted as a cloud of cosmic strings with spherical symmetry \cite{9}.
Hence, the spacetime is conical. However, with the inclusion of a linear or
nonlinear electromagnetic field, the spacetime is no more conical in the
context of $f(R)$ gravity.

Within the framework of ETG gravity, black hole solutions have been widely
studied in the literature (see \cite{1,10} and references therein for a
complete review). However, the solutions that result with naked
singularities have not been studied in detail. In physics, naked
singularities are considered to be a threat to the cosmic censorship
hypothesis. Furthermore, as in classical general relativity, compared to the
black hole solutions, naked singularities are not well understood in the
context of $f(R)$ gravity. This still remains a fundamental problem in
general relativity as well as in ETG to be solved. Another important diff\i
culty in resolving this problem is the scale on which the curvature
singularity occurs. On these small scales, it is believed that the classical
methods should be replaced with quantum techniques in resolving the
singularity problems that necessitate the use of quantum gravity. Since the
quantum theory of gravity is still "under construction", an alternative
method is proposed by Wald \cite{11} which was further developed by Horowitz
and Marolf (HM) \cite{12} in determining the character of classically
singular spacetime and to see if quantum effects have any chance to heal or
regularize the dynamics and restore the predictability if the singularity is
probed with quantum particles/fields.

In this paper, we investigate the occurrence of naked singularities in the
context of $f(R)$ gravity from the point of view of quantum mechanics. We
believe that this will be the unique example wherein the formation of a
classically naked curvature singularities in $f(R)$ gravity will be probed
with quantum fields/particles that obey the Klein$-$Gordon, Dirac and
Maxwell equations. The criterion proposed by HM will be used in this study
to investigate the occurrence of naked singularities.

This criterion has been used successfully for other spacetimes to check
whether the classically singular spacetimes are quantum mechanically regular
or not. As an example; negative mass Schwarzschild spacetime, charged
dilatonic black hole spacetime and fundamental string spacetimes are
considered in \cite{12}. An alternative function space, namely the Sobelov
space instead of the Hilbert space, has been introduced in \cite{13}, for
analyzing the singularities within the framework of quantum mechanics.
Helliwell and Konkowski have studied quasiregular \cite{14}, Gal'tsov$-$%
Letelier$-$Tod spacetime \cite{15}, Levi-Civita spacetimes \cite{16,17}, and
recently, they have also considered conformally static spacetimes \cite{18}.
Pitelli and Letelier have studied spherical and cylindrical topological
defects \cite{19}, Banados$-$Teitelboim$-$Zanelli (BTZ) spacetimes \cite{20}%
, the global monopole spacetime \cite{21} and cosmological spacetimes \cite%
{22}. Quantum singularities in matter coupled $2+1$ dimensional black hole
spacetimes are considered in \cite{23}. Quantum singularities are also
considered in Lovelock theory \cite{24} and linear dilaton black hole
spacetimes \cite{25}. Recently, the occurrence of naked singularities in a $%
2+1$ dimensional magnetically charged solution in Einstein$-$Power$-$Maxwell
theory have also been considered \cite{26}.

The main theme in these studies is to understand whether these classically
singular spacetimes turn out to be quantum mechanically regular if they are 
probed with quantum fields rather than classical particles.

The solution to be investigated in this paper is a kind of $f(R)$ gravity
extension of the analysis presented in \cite{21} for the global monopole
spacetime. The inclusion of the linear Maxwell field within the context of $%
f(R)$ gravity \ affects the topology significantly and removes the conical
nature at infinity. Furthermore, the true timelike naked curvature
singularity is created at $r=0$ which is peculiar to spherically symmetric
systems. We investigate this singularity within the framework of quantum
mechanics by employing three different quantum fields/particles obeying the
Klein$-$Gordon, Dirac and Maxwell fields with different spin structures.

The paper is organized as follows: In Sec.II, we review the solution found
recently in \cite{7}, and give the structure of the spacetime. In Sec.III,
first, the definition of quantum singularity for static spacetimes is
briefly introduced. Then, the quantum fields obeying the Klein$-$Gordon,
Dirac and Maxwell equations are used to probe the singularity. The paper
ends with a conclusion in Sec. IV. \ \ \ 

\section{The Metric for $f(R)$ Gravity Coupled to Maxwell Fields and
Spacetime Structure}

Recently, an exact analytic solution for $f\left( R\right) $ gravity coupled
with linear and nonlinear Maxwell field in four dimensions has been
presented in \cite{7}. The corresponding action for $f\left( R\right) $
gravity coupled with linear Maxwell field in four dimensions is given by,%
\begin{equation}
S=\int d^{4}x\sqrt{-g}\left[ \frac{f\left( R\right) }{2\kappa }-\frac{1}{%
4\pi }F\right] ,
\end{equation}%
in which $f\left( R\right) $ is a real function of the Ricci scalar $R,$ and 
$F=\frac{1}{4}F_{\mu \nu }F^{\mu \nu }$ is the Maxwell invariant. The
Maxwell two-form is given by%
\begin{equation}
\mathbf{F}=\frac{Q}{r^{2}}dt\wedge dr+P\sin \theta d\theta \wedge d\varphi ,
\end{equation}%
in which $Q$ and $P$ are the electric and magnetic charges, respectively.
The static spherically symmetric metric ansatz is%
\begin{equation}
ds^{2}=-B\left( r\right) dt^{2}+\frac{dr^{2}}{B\left( r\right) }+r^{2}\left(
d\theta ^{2}+\sin ^{2}\theta d\varphi ^{2}\right) ,
\end{equation}%
where $B\left( r\right) $ stands for the only metric function to be found.
The Maxwell equations $\left( \text{i.e. }dF=0=d^{\ast }F\right) $\ are
satisfied and the field equations are given by%
\begin{equation}
f_{R}R_{\mu }^{\nu }+\left( \square f_{R}-\frac{1}{2}f\right) \delta _{\mu
}^{\nu }-\nabla ^{\nu }\nabla _{\mu }f_{R}=\kappa T_{\mu }^{\nu },
\end{equation}%
in which 
\begin{eqnarray}
f_{R} &=&\frac{df\left( R\right) }{dR}, \\
\square f_{R} &=&\frac{1}{\sqrt{-g}}\partial _{\mu }\left( \sqrt{-g}\partial
^{\mu }\right) f_{R}, \\
\nabla ^{\nu }\nabla _{\mu }f_{R} &=&g^{\alpha \nu }\left[ \left(
f_{R}\right) _{,\mu ,\alpha }-\Gamma _{\mu \alpha }^{m}\left( f_{R}\right)
_{,m}\right] ,
\end{eqnarray}%
while the energy momentum tensor is 
\begin{equation}
4\pi T_{\mu }^{\nu }=-F\delta _{\mu }^{\nu }+F_{\mu \lambda }F^{\nu \lambda
}.
\end{equation}%
Furthermore, the trace of the field equation (4) reads%
\begin{equation}
f_{R}R+\left( d-1\right) \square f_{R}-\frac{d}{2}f=\kappa T,
\end{equation}%
with $T=T_{\mu }^{\mu }.$ The non-zero energy momentum tensor components are%
\begin{equation}
T_{\mu }^{\nu }=\frac{P^{2}+Q^{2}}{8\pi r^{4}}diag\left[ -1,-1,1,1\right] ,
\end{equation}%
and with zero trace, we have%
\begin{equation}
f=\frac{1}{2}f_{R}R+3\square f_{R}.
\end{equation}%
With reference to the paper \cite{7}, the form of the function $f\left(
R\right) $ is assumed to be ,

\begin{equation}
f\left( R\right) =\xi \left( R+\frac{1}{2}R_{0}\right) +2\alpha \sqrt{R+R_{0}%
},
\end{equation}%
which leads to 
\begin{equation}
R=\frac{\alpha ^{2}}{\eta ^{2}r^{2}}-R_{0},
\end{equation}%
where $\alpha $ , $R_{0},$ and $\xi $ are constants. Consequently, the
metric function $B(r)$ is obtained for the free parameters $\alpha =\eta $
as, 
\begin{equation}
B\left( r\right) =\frac{1}{2}-\frac{m}{r}+\frac{q^{2}}{r^{2}}-\frac{\Lambda
_{eff}}{3}r^{2},
\end{equation}%
where $m=\frac{-\xi }{3\eta },$ $\Lambda _{eff}=\frac{-R_{0}}{4}$ and $q^{2}=%
\frac{Q^{2}+P^{2}}{\xi }.$ As was explained in \cite{7}, due to the
constraints on the free parameters, this solution does not admit the Reissner%
$-$Nordstr\"{o}m (RN)$-$de Sitter (dS) limit. However, in the limit $\xi =1$
and $P=Q=0,$ the solution reduces to the well known global monopole solution
reported in \cite{8}, which represents a spherically symmetric,
non-asymptotically flat solution with a matter field that extends to
infinity.\ Furthermore, this solution can also be considered as a
spherically symmetric cloud of cosmic string which gives rise to a deficit
angle \cite{9}. Therefore, the solution given in equation (14) , is a kind
of Einstein$-$Maxwell extension of the global monopole solution in $f\left(
R\right) $ gravity. One of the striking effects of the additional fields is
the removal of the conical geometry of the global monopole spacetime. The
Kretschmann scalar which indicates the formation of curvature singularity is
given by%
\begin{equation*}
\mathcal{K}=\frac{1}{3}\frac{8\lambda ^{2}r^{8}+4\lambda
r^{6}+3r^{4}+12mr^{3}+12r^{2}\left( 3m^{2}-q^{2}\right) -144mq^{2}r++168q^{4}%
}{r^{8}}.
\end{equation*}%
It is obvious that $r=0$ is a typical central curvature singularity. This is
a timelike naked singularity because the behavior of the new radial
coordinate defined by $r_{\ast }=\int \frac{dr}{B(r)}$ is finite when $%
r\rightarrow 0.$ Hence, the new solution obtained in \cite{7} and given in
equation (14) is classically a singular spacetime.

Our aim in the next section is to investigate this classically singular
spacetime with regard to the quantum mechanical point of view.

\section{Quantum Singularities}

One of the important predictions of the Einstein's theory of general
relativity is the formation of spacetime singularities. In classical general
relativity, singularities are defined as the points in which the evolution
of timelike or null geodesics is not defined after a proper time. According
to the classification of the classical singularities devised by Ellis and
Schmidt , scalar curvature singularities are the strongest ones in the sense
that the spacetime cannot be extended and all physical quantities, such as
the gravitational field, energy density and tidal forces, diverge at the
singular point. In black hole spacetimes, the location of the curvature
singularity is at $r=0$ and is covered by horizon(s). As long as the
singularities are hidden by horizon(s), they do not constitute a threat to
the Penrose cosmic censorship hypothesis. However, there are some cases that
the singularity is not hidden and hence, it is \textit{naked}. In the case
of naked singularities, further care is required because they violate the
cosmic censorship hypothesis. The resolution of the naked singularities
stands as one of the most drastic problems in general relativity to be
solved.

Naked singularities that occur at $r=0$ are on the very small scales where
classical general relativity is expected to be replaced by quantum theory of
gravity. In this paper, the occurrence of naked singularities in $f(R)$
gravity will be analyzed through a quantum mechanical point of view. In
probing the singularity, quantum test particles/fields obeying the Klein$-$%
Gordon, Dirac and Maxwell equations are used. In other words, the
singularity will be probed with spin $0$, spin $1/2$ and spin $1$ fields.
The reason for using three different types of field is to clarify whether or
not the classical singularity is sensitive to the spin of the fields.

Our analysis will be based on the pioneering work of Wald, which was further
developed by HM to probe the classical singularities with quantum test
particles obeying the Klein$-$Gordon equation in static spacetimes having
timelike singularities. According to HM, the singular character of the
spacetime is defined as the ambiguity in the evolution of the wave
functions. That is to say, the singular character is determined in terms of
the ambiguity when attempting to find a self-adjoint extension of the
operator to the entire Hilbert space. If the extension is unique, it is said
that the space is quantum mechanically regular. A brief review now follows:

Consider a static spacetime $\left( M,g_{\mu \nu }\right) $\ with a timelike
Killing vector field $\xi ^{\mu }$. Let $t$ denote the Killing parameter and 
$\Sigma $\ denote a static slice. The Klein$-$Gordon equation in this space
is

\begin{equation}
\left( \nabla ^{\mu }\nabla _{\mu }-M^{2}\right) \psi =0.
\end{equation}%
This equation can be written in the form

\begin{equation}
\frac{\partial ^{2}\psi }{\partial t^{2}}=\sqrt{f}D^{i}\left( \sqrt{f}%
D_{i}\psi \right) -fM^{2}\psi =-A\psi ,
\end{equation}%
in which $f=-\xi ^{\mu }\xi _{\mu }$ and $D_{i}$ is the spatial covariant
derivative on $\Sigma $. The Hilbert space $\mathcal{H}$, $\left(
L^{2}\left( \Sigma \right) \right) $\ is the space of square integrable
functions on $\Sigma $. The domain of an operator $A,$ $D(A),$ is taken in
such a way that it does not enclose the spacetime singularities. An
appropriate set is $C_{0}^{\infty }\left( \Sigma \right) $, the set of
smooth functions with compact support on $\Sigma $. The operator $A$ is
real, positive and symmetric; therefore, its self-adjoint extensions always
exist. If \ it has a unique extension $A_{E},$ then $A$ is called
essentially self-adjoint \cite{27,28,29}. Accordingly, the Klein$-$Gordon
equation for a free particle satisfies

\begin{equation}
i\frac{d\psi}{dt}=\sqrt{A_{E}}\psi,
\end{equation}
with the solution

\begin{equation}
\psi \left( t\right) =\exp \left[ -it\sqrt{A_{E}}\right] \psi \left(
0\right) .
\end{equation}%
If $A$ is not essentially self-adjoint, the future time evolution of the
wave function (18) is ambiguous. Then the HM criterion defines the spacetime
as quantum mechanically singular. However, if there is only a single
self-adjoint extension, the operator $A$ is said to be\ essentially
self-adjoint and the quantum evolution described by Eq.(18) is uniquely
determined by the initial conditions. According to the HM criterion, this
spacetime is said to be quantum mechanically non-singular. In order to
determine the number of self-adjoint extensions, the concept of deficiency
indices is used. The deficiency subspaces $N_{\pm }$ are defined by ( see
Ref. \cite{13} for a detailed mathematical background) 

\begin{align}
N_{+}& =\{\psi \in D(A^{\ast }),\text{ \ \ \ \ \ \ }A^{\ast }\psi =Z_{+}\psi
,\text{ \ \ \ \ \ }ImZ_{+}>0\}\text{ \ \ \ \ \ with dimension }n_{+} \\
N_{-}& =\{\psi \in D(A^{\ast }),\text{ \ \ \ \ \ \ }A^{\ast }\psi =Z_{-}\psi
,\text{ \ \ \ \ \ }ImZ_{-}<0\}\text{ \ \ \ \ \ with dimension }n_{-}  \notag
\end{align}%
The dimensions $\left( \text{ }n_{+},n_{-}\right) $ are the deficiency
indices of the operator $A$. The indices $n_{+}(n_{-})$ are completely
independent of the choice of $Z_{+}(Z_{-})$ depending only on whether or not 
$Z$ lies in the upper (lower) half complex plane. Generally one takes $%
Z_{+}=i\lambda $ and $Z_{-}=-i\lambda $ , where $\lambda $ is an arbitrary
positive constant necessary for dimensional reasons. The determination of
deficiency indices is then reduced to counting the number of solutions of $%
A^{\ast }\psi =Z\psi $ ; (for $\lambda =1$),

\begin{equation}
A^{\ast }\psi \pm i\psi =0
\end{equation}%
that belong to the Hilbert space $\mathcal{H}$. If there are no square
integrable solutions ( i.e. $n_{+}=n_{-}=0)$, the operator $A$ possesses a
unique self-adjoint extension and  essentially self-adjoint. Consequently,
the way to find a sufficient condition for the operator $A$ to be
essentially self-adjoint is to investigate the solutions satisfying Eq. (20)
that do not belong to the Hilbert space.

\subsection{Klein$-$Gordon Fields}

The Klein$-$Gordon equation for a scalar particle with mass $M$ is given by

\begin{equation}
\square \psi =g^{-1/2}\partial _{\mu }\left[ g^{1/2}g^{\mu \nu }\partial
_{\nu }\right] \psi =M^{2}\psi .
\end{equation}%
For the metric (3), the Klein$-$Gordon equation becomes

\begin{eqnarray}
\frac{\partial ^{2}\psi }{\partial t^{2}} &=&-B\left( r\right) \left\{
B\left( r\right) \frac{\partial ^{2}\psi }{\partial r^{2}}+\frac{1}{r^{2}}%
\frac{\partial ^{2}\psi }{\partial \theta ^{2}}+\frac{1}{r^{2}\sin
^{2}\theta }\frac{\partial ^{2}\psi }{\partial \varphi ^{2}}+\frac{\cot
\theta }{r^{2}}\frac{\partial \psi }{\partial \theta }+\left( \frac{2B\left(
r\right) }{r}+B^{^{\prime }}\left( r\right) \right) \frac{\partial \psi }{%
\partial r}\right\}  \\
&&+B\left( r\right) M^{2}\psi .  \notag
\end{eqnarray}%
In analogy with  equation (16), the spatial operator $A$ for the massless
case is

\begin{equation}
\emph{A}=B\left( r\right) \left\{ B\left( r\right) \frac{\partial ^{2}}{%
\partial r^{2}}+\frac{1}{r^{2}}\frac{\partial ^{2}}{\partial \theta ^{2}}+%
\frac{1}{r^{2}\sin ^{2}\theta }\frac{\partial ^{2}}{\partial \varphi ^{2}}+%
\frac{\cot \theta }{r^{2}}\frac{\partial }{\partial \theta }+\left( \frac{%
2B\left( r\right) }{r}+B^{^{\prime }}\left( r\right) \right) \frac{\partial 
}{\partial r}\right\} ,
\end{equation}%
and the equation to be solved is $\left( \emph{A}^{\ast }\pm i\right) \psi
=0.$Using separation of variables, $\psi =R\left( r\right) Y_{l}^{m}\left(
\theta ,\varphi \right) $, we get the radial portion of equation (20) as

\begin{equation}
\frac{d^{2}R\left( r\right) }{dr^{2}}+\frac{\left( r^{2}B\left( r\right)
\right) ^{^{\prime }}}{r^{2}B\left( r\right) }\frac{dR\left( r\right) }{dr}%
+\left( \frac{-l\left( l+1\right) }{r^{2}B\left( r\right) }\pm \frac{i}{%
B^{2}\left( r\right) }\right) R\left( r\right) =0.
\end{equation}%
where a prime denotes the derivative with respect to $r$.

\subsubsection{The case of r$\rightarrow \infty $}

The case $r\rightarrow \infty $ is topologically different compared to the
analysis reported in \cite{21}. In the present problem the geometry is not
conical. The approximate metric when $r\rightarrow \infty $ is%
\begin{equation}
ds^{2}\simeq -(\frac{R_{0}r^{2}}{12})dt^{2}+\left( \frac{12}{R_{0}r^{2}}%
\right) dr^{2}+r^{2}\left( d\theta ^{2}+\sin ^{2}\theta d\varphi ^{2}\right)
.
\end{equation}%
For the above metric, the radial equation (24) becomes,%
\begin{equation}
\frac{d^{2}R\left( r\right) }{dr^{2}}+\frac{4}{r}\frac{dR\left( r\right) }{dr%
}=0,
\end{equation}%
whose solution is%
\begin{equation*}
R\left( r\right) =C_{1}+\frac{C_{2}}{r^{3}},
\end{equation*}%
where $C_{1}$ and $C_{2}$ are arbitrary integration constants. It is clearly
observed that the above solution is square integrable as $r\rightarrow
\infty $ if and only if $C_{1}=0.$ Hence, the asymptotic behavior of $R(r)$
is given by $R(r)\simeq \frac{C_{2}}{r^{3}}.$

\subsubsection{The case of r$\rightarrow 0$}

Near the origin there is a true timelike curvature singularity resulting
from the existence of charge. Therefore, the approximate metric near the
origin is given by%
\begin{equation}
ds^{2}\simeq -(\frac{q^{2}}{r^{2}})dt^{2}+\left( \frac{r^{2}}{q^{2}}\right)
dr^{2}+r^{2}\left( d\theta ^{2}+\sin ^{2}\theta d\varphi ^{2}\right) .
\end{equation}

The radial equation (24) for the above metric reduces to

\begin{equation}
\frac{d^{2}R\left( r\right) }{dr^{2}}-\frac{l\left( l+1\right) }{q^{2}}%
R\left( r\right) =0,
\end{equation}%
whose solution is

\begin{eqnarray}
R\left( r\right) &=&C_{3}e^{\alpha r}+C_{4}e^{-\alpha r} \\
\alpha &=&\frac{\sqrt{l\left( l+1\right) }}{q}  \notag
\end{eqnarray}%
where $C_{3}$\ and $C_{4}$ are arbitrary integration constants. The square
integrability of the above solution is checked by calculating the squared
norm of the above solution in which the function space on each $t=$ constant
hypersurface $\Sigma $ is defined as $\mathcal{H=}\{R\mid \parallel
R\parallel <\infty \}.$ The squared norm for the metric (27) is given by,

\begin{equation}
\parallel R\parallel ^{2}=\int_{0}^{\text{constant}}\frac{\left\vert R\left(
r\right) \right\vert ^{2}r^{4}}{q^{2}}dr.
\end{equation}%
Our calculation has revealed that the solution above is always square
integrable near $r=0,$ even if $l=0,$ which corresponds to the $S$-wave
solutions. \ 

Consequently, the spatial operator $A$ has deficiency indices $n_{+}=n_{-}=1,
$ and it is not essentially self-adjoint. Hence, the classical singularity
at $r=0$ remains quantum mechanically singular when probed with fields
obeying the Klein$-$Gordon equation.

\subsection{Maxwell fields}

The Newman$-$Penrose formalism will be used to find the source-free Maxwell
fields propagating in the space of $f(R)$ gravity. Let us note that the
signature of the metric (3) is changed to $-2$ in order to use the
source-free Maxwell equations in Newman$-$Penrose formalism. Thus, the
metric (3) is rewritten as,%
\begin{equation}
ds^{2}=B\left( r\right) dt^{2}-\frac{dr^{2}}{B\left( r\right) }-r^{2}\left(
d\theta ^{2}+\sin ^{2}\theta d\varphi ^{2}\right) .
\end{equation}%
The four coupled source-free Maxwell equations for electromagnetic fields in
the Newman$-$Penrose formalism is given by%
\begin{eqnarray}
D\phi _{1}-\bar{\delta}\phi _{0} &=&\left( \pi -2\alpha \right) \phi
_{0}+2\rho \phi _{1}-\kappa \phi _{2}, \\
\delta \phi _{2}-\Delta \phi _{1} &=&-\nu \phi _{0}+2\mu \phi _{1}+\left(
\tau -2\beta \right) \phi _{2},  \notag \\
\delta \phi _{1}-\Delta \phi _{0} &=&\left( \mu -2\gamma \right) \phi
_{0}+2\tau \phi _{1}-\sigma \phi _{2},  \notag \\
D\phi _{2}-\bar{\delta}\phi _{1} &=&-\lambda \phi _{0}+2\pi \phi _{1}+\left(
\rho -2\epsilon \right) \phi _{2},  \notag
\end{eqnarray}%
where $B(r)$ is the metric function given in Eq.(14), $\phi _{0},$ $\phi _{1}
$ and $\phi _{2}$ are the Maxwell spinors, $\epsilon ,\rho ,\pi ,\alpha ,\mu
,\gamma ,\beta $ and $\tau $ are the spin coefficients to be found and the
bar denotes  complex conjugation. The null tetrad vectors for the metric
(31) are defined by%
\begin{eqnarray}
l^{a} &=&\left( \frac{1}{B(r)},1,0,0\right) , \\
n^{a} &=&\left( \frac{1}{2},-\frac{B(r)}{2},0,0\right) ,  \notag \\
m^{a} &=&\frac{1}{\sqrt{2}}\left( 0,0,\frac{1}{r},\frac{i}{r\sin \theta }%
\right) .  \notag
\end{eqnarray}%
The directional derivatives in the Maxwell's equations are defined by $%
D=l^{a}\partial _{a},\Delta =n^{a}\partial _{a}$ and $\delta =m^{a}\partial
_{a}.$ We define operators in the following way

\begin{eqnarray}
\mathbf{D}_{0} &=&D,  \notag \\
\mathbf{D}_{0}^{\dagger } &=&-\frac{2}{B\left( r\right) }\Delta , \\
\mathbf{L}_{0}^{\dagger } &=&\sqrt{2}r\text{ }\delta \text{ and }\mathbf{L}%
_{1}^{\dagger }=\mathbf{L}_{0}^{\dagger }+\frac{\cot \theta }{2},  \notag \\
\mathbf{L}_{0} &=&\sqrt{2}r\text{ }\bar{\delta}\text{ and }\mathbf{L}_{1}=%
\mathbf{L}_{0}+\frac{\cot \theta }{2}.  \notag
\end{eqnarray}%
The non-zero spin coefficients are 
\begin{equation}
\mu =-\frac{1}{r}\frac{B(r)}{2},\text{ \ \ \ \ }\rho =-\frac{1}{r},\text{ \
\ \ }\gamma =\frac{1}{4}B^{^{\prime }}(r),\text{ \ \ \ \ }\beta =-\alpha =%
\frac{1}{2\sqrt{2}}\frac{\cot \theta }{r}.
\end{equation}%
The Maxwell spinors are defined by \cite{30} 
\begin{eqnarray}
\phi _{0} &=&F_{13}=F_{\mu \nu }l^{\mu }m^{\nu } \\
\phi _{1} &=&\frac{1}{2}\left( F_{12}+F_{43}\right) =\frac{1}{2}F_{\mu \nu
}\left( l^{\mu }n^{\nu }+\overline{m}^{\mu }m^{\nu }\right) ,  \notag \\
\phi _{2} &=&F_{42}=F_{\mu \nu }\overline{m}^{\mu }n^{\nu },  \notag
\end{eqnarray}%
where $F_{ij}\left( i,j=1,2,3,4\right) $ and $F_{\mu \nu }\left( \mu ,\nu
=0,1,2,3\right) $ are the components of the Maxwell tensor in the tetrad and
tensor bases, respectively. Substituting Eq.(34) into the Maxwell's
equations together with non-zero spin coefficients, the Maxwell equations
become

\begin{gather}
\left( \mathbf{D}_{0}+\frac{2}{r}\right) \phi _{1}-\frac{1}{r\sqrt{2}}%
\mathbf{L}_{1}\phi _{0}=0, \\
\left( \mathbf{D}_{0}+\frac{1}{r}\right) \phi _{2}-\frac{1}{r\sqrt{2}}%
\mathbf{L}_{0}\phi _{1}=0, \\
\frac{B\left( r\right) }{2}\left( \mathbf{D}_{0}^{\dagger }+\frac{%
B^{^{\prime }}\left( r\right) }{B\left( r\right) }+\frac{1}{r}\right) \phi
_{0}+\frac{1}{r\sqrt{2}}\mathbf{L}_{0}^{\dagger }\phi _{1}=0, \\
\frac{B\left( r\right) }{2}\left( \mathbf{D}_{0}^{\dagger }+\frac{2}{r}%
\right) \phi _{1}+\frac{1}{r\sqrt{2}}\mathbf{L}_{1}^{\dagger }\phi _{2}=0.
\end{gather}%
The equations\ above will become more tractable if the variables are changed
to

\begin{equation*}
\Phi _{0}=\phi _{0}e^{ikt},\text{ \ \ }\Phi _{1}=\sqrt{2}r\phi _{1}e^{ikt},%
\text{ \ \ \ \ }\Phi _{2}=2r^{2}\phi _{2}e^{ikt}.
\end{equation*}%
Then we have%
\begin{gather}
\left( \mathbf{D}_{0}+\frac{1}{r}\right) \Phi _{1}-\mathbf{L}_{1}\Phi _{0}=0,
\\
\left( \mathbf{D}_{0}-\frac{1}{r}\right) \Phi _{2}-\mathbf{L}_{0}\Phi _{1}=0,
\\
r^{2}B\left( r\right) \left( \mathbf{D}_{0}^{\dagger }+\frac{B^{^{\prime
}}\left( r\right) }{B\left( r\right) }+\frac{1}{r}\right) \Phi _{0}+\mathbf{L%
}_{0}^{\dagger }\Phi _{1}=0, \\
r^{2}B\left( r\right) \left( \mathbf{D}_{0}^{\dagger }+\frac{1}{r}\right)
\Phi _{1}+\mathbf{L}_{1}^{\dagger }\Phi _{2}=0.
\end{gather}%
The commutativity of the operators $\mathbf{L}$ and $\mathbf{D}$ enables us
to eliminate each $\Phi _{i}$ from above equations, and hence we have 
\begin{gather}
\left[ \mathbf{L}_{0}^{\dagger }\mathbf{L}_{1}+r^{2}B\left( r\right) \left( 
\mathbf{D}_{0}+\frac{B^{^{\prime }}\left( r\right) }{B\left( r\right) }+%
\frac{3}{r}\right) \left( \mathbf{D}_{0}^{\dagger }+\frac{B^{^{\prime
}}\left( r\right) }{B\left( r\right) }+\frac{1}{r}\right) \right] \Phi
_{0}\left( r,\theta \right) =0, \\
\left[ \mathbf{L}_{0}\mathbf{L}_{1}^{\dagger }+r^{2}B\left( r\right) \left( 
\mathbf{D}_{0}^{\dagger }+\frac{1}{r}\right) \left( \mathbf{D}_{0}-\frac{1}{r%
}\right) \right] \Phi _{2}\left( r,\theta \right) =0, \\
\left[ \mathbf{L}_{1}\mathbf{L}_{0}^{\dagger }+r^{2}B\left( r\right) \left( 
\mathbf{D}_{0}^{\dagger }+\frac{B^{^{\prime }}\left( r\right) }{B\left(
r\right) }+\frac{1}{r}\right) \left( \mathbf{D}_{0}+\frac{1}{r}\right) %
\right] \Phi _{1}\left( r,\theta \right) =0.
\end{gather}%
The variables $r$ and $\theta $ can be separated by assuming a separable
solution in the form of%
\begin{equation*}
\Phi _{0}\left( r,\theta \right) =f_{0}\left( r\right) \Theta _{0}\left(
\theta \right) ,\text{ \ \ }\Phi _{1}\left( r,\theta \right) =f_{1}\left(
r\right) \Theta _{1}\left( \theta \right) ,\text{ \ \ \ \ }\Phi _{2}\left(
r,\theta \right) =f_{2}\left( r\right) \Theta _{2}\left( \theta \right) .
\end{equation*}%
The separation constants for Eq. (45) and Eq. (46) are the same, because $%
\mathbf{L}_{n}=-\mathbf{L}_{n}^{\dagger }\left( \pi -\theta \right) ,$ or,
in other words, the operator $\mathbf{L}_{0}^{\dagger }\mathbf{L}_{1}$
acting on $\Theta _{0}\left( \theta \right) $ is the same as the operator $%
\mathbf{L}_{0}\mathbf{L}_{1}^{\dagger }$ acting on $\Theta _{2}\left( \theta
\right) $ if we replace $\theta $ by $\pi -\theta $. However, for Eq. (47)
we will assume another separation constant. Furthermore, by defining $%
R_{0}\left( r\right) =\frac{f_{0}(r)}{rB\left( r\right) }$, $R_{1}(r)=\frac{%
f_{1}\left( r\right) }{r}$ and $R_{2}(r)=\frac{f_{2}\left( r\right) }{r}$,
the radial equations can be written as 
\begin{gather}
f_{0}^{^{\prime \prime }}(r)+\frac{2}{r}f_{0}^{^{\prime }}(r)+ \\
\left[ -i\omega \left( \frac{2}{rB\left( r\right) }-\frac{B^{^{\prime
}}\left( r\right) }{B^{2}\left( r\right) }\right) +\frac{\omega ^{2}}{%
B^{2}\left( r\right) }-\frac{\epsilon ^{2}}{r^{2}B\left( r\right) }\right]
f_{0}(r)=0,  \notag
\end{gather}%
\begin{gather}
f_{2}^{^{\prime \prime }}(r)-\frac{2}{r}f_{2}^{^{\prime }}(r)+ \\
\left[ i\omega \left( \frac{2}{rB\left( r\right) }-\frac{B^{^{\prime
}}\left( r\right) }{B^{2}\left( r\right) }\right) +\frac{\omega ^{2}}{%
B^{2}\left( r\right) }-\frac{\epsilon ^{2}}{r^{2}B\left( r\right) }\right]
f_{2}(r)=0,  \notag
\end{gather}%
\begin{gather}
f_{1}^{^{\prime \prime }}(r)+\frac{B^{^{\prime }}\left( r\right) }{B\left(
r\right) }f_{1}^{^{\prime }}(r)+ \\
\left[ \frac{\omega ^{2}}{B^{2}\left( r\right) }-\frac{\eta ^{2}}{%
r^{2}B\left( r\right) }\right] f_{1}(r)=0,  \notag
\end{gather}%
where $\epsilon $ and $\eta $ are the separability constants.

\subsubsection{The case r$\rightarrow \infty $}

For the case $r\rightarrow \infty $, the corresponding metric is given in
Eq.(25). Hence, the radial parts of the Maxwell equations, (48) , (49) and
(50), become%
\begin{eqnarray}
f_{j}^{^{\prime \prime }}(r)+\frac{2}{r}f_{j}^{^{\prime }}(r) &=&0,\text{ \
\ \ \ \ }j=0,1\text{\ \ \ \ \ } \\
\text{\ }f_{2}^{^{\prime \prime }}(r)-\frac{2}{r}f_{2}^{^{\prime }}(r) &=&0
\end{eqnarray}%
Thus, the solutions in the asymptotic case are%
\begin{eqnarray}
R_{j}(r) &=&C_{1}+\frac{C_{2}}{r},\text{ \ \ \ }j=0,1\text{\ \ } \\
R_{2}(r) &=&C_{3}+\frac{C_{4}}{r^{3}},
\end{eqnarray}%
in which $C_{i}$ are integration constants. The solution above is square
integrable if $C_{1}=$ $C_{3}=0.$ Therefore, the asymptotic form of the
solutions behaves as $R_{j}(r)\sim \frac{C_{2}}{r},$ \ \ \ $j=0,1$ and $%
R_{2}(r)\sim \frac{C_{4}}{r^{3}}.$

\subsubsection{The case r$\rightarrow 0$}

The metric near $r\rightarrow 0$ is given in Eq.(27). Hence, the radial
parts of the Maxwell equations (48) , (49) and (50) for this case are given
by%
\begin{eqnarray}
R_{j}^{^{\prime \prime }}(r)-\frac{2}{r}R_{j}^{^{\prime }}(r)-\frac{\alpha
^{2}}{q^{2}}R_{j}(r) &=&0\text{ , \ }j=1,2 \\
R_{0}^{^{\prime \prime }}(r)+\frac{2}{r}R_{0}^{^{\prime }}(r)-\frac{\eta ^{2}%
}{q^{2}}R_{0}(r) &=&0\text{\ }
\end{eqnarray}%
whose solutions are obtained as,%
\begin{eqnarray}
R_{j}(r) &=&C_{3}e^{\frac{\alpha }{q}r}\left( \alpha r-1\right) +C_{4}e^{-%
\frac{\alpha }{q}r}\left( \alpha r+1\right) ,\text{ \ \ \ \ \ }j=1,2, \\
R_{0}(r) &=&\frac{C_{5}}{r}\sinh \left( \frac{\eta }{q}r\right) +\frac{C_{6}%
}{r}\cosh \left( \frac{\eta }{q}r\right) 
\end{eqnarray}%
where $C_{i}$ are constants. The above solution is checked for square
integrability. Calculations have revealed that 
\begin{equation*}
\parallel R_{i}\parallel ^{2}=\int_{0}^{\text{constant}}\frac{\left\vert
R_{i}\left( r\right) \right\vert ^{2}r^{4}}{q^{2}}dr<\infty ,
\end{equation*}%
which indicates that the obtained solutions are square integrable. The
definition of the quantum singularity for Maxwell fields will be the same as
for the Klein$-$Gordon fields. Here, since we have three equations governing
the dynamics of the photon waves, the unique self-adjoint extension
condition on the spatial part of the Maxwell operator should be examined for
each of the three equations. As a result, the occurrence of the naked
singularity in $f(R)$ gravity is quantum mechanically singular if it is
probed with photon waves.

\subsection{Dirac Fields}

The Newman$-$Penrose formalism will also be used here to find the massless
Dirac fields (fermions) propagating in the space of $f(R)$-gravity. The
Chandrasekhar-Dirac (CD) equations in the Newman$-$Penrose formalism are
given by

\begin{eqnarray}
\left( D+\epsilon -\rho \right) F_{1}+\left( \bar{\delta}+\pi -\alpha
\right) F_{2} &=&0, \\
\left( \Delta +\mu -\gamma \right) F_{2}+\left( \delta +\beta -\tau \right)
F_{1} &=&0,  \notag \\
\left( D+\bar{\epsilon}-\bar{\rho}\right) G_{2}-\left( \delta +\bar{\pi}-%
\bar{\alpha}\right) G_{1} &=&0,  \notag \\
\left( \Delta +\bar{\mu}-\bar{\gamma}\right) G_{1}-\left( \bar{\delta}+\bar{%
\beta}-\bar{\tau}\right) G_{2} &=&0,  \notag
\end{eqnarray}%
where $F_{1},F_{2},G_{1}$ and $G_{2}$ are the components of the wave
function, $\epsilon ,\rho ,\pi ,\alpha ,\mu ,\gamma ,\beta $ and $\tau $ are
the spin coefficients to be found. The non-zero spin coefficients are given
in Eq.(35). The directional derivatives in the CD equations are the same as
in the Maxwell equations. Substituting non-zero spin coefficients and the
definitions of the operators given in Eq.(34) into the CD equations leads to

\begin{gather}
\left( \mathbf{D}_{0}+\frac{1}{r}\right) F_{1}+\frac{1}{r\sqrt{2}}\mathbf{L}%
_{1}F_{2}=0,  \notag \\
-\frac{B\left( r\right) }{2}\left( \mathbf{D}_{0}^{\dagger }+\frac{%
B^{^{\prime }}\left( r\right) }{2B\left( r\right) }+\frac{1}{r}\right) F_{2}+%
\frac{1}{r\sqrt{2}}\mathbf{L}_{1}^{\dagger }F_{1}=0,  \notag \\
\left( \mathbf{D}_{0}+\frac{1}{r}\right) G_{2}-\frac{1}{r\sqrt{2}}\mathbf{L}%
_{1}^{\dagger }G_{1}=0,  \notag \\
\frac{B\left( r\right) }{2}\left( \mathbf{D}_{0}^{\dagger }+\frac{%
B^{^{\prime }}\left( r\right) }{2B\left( r\right) }+\frac{1}{r}\right) G_{1}+%
\frac{1}{r\sqrt{2}}\mathbf{L}_{1}G_{2}=0.
\end{gather}%
For the solution of the CD equations, we assume a separable solution in the
form of%
\begin{eqnarray}
F_{1} &=&f_{1}(r)Y_{1}(\theta )e^{i\left( kt+m\varphi \right) }, \\
F_{2} &=&f_{2}(r)Y_{2}(\theta )e^{i\left( kt+m\varphi \right) },  \notag \\
G_{1} &=&g_{1}(r)Y_{3}(\theta )e^{i\left( kt+m\varphi \right) },  \notag \\
G_{2} &=&g_{2}(r)Y_{4}(\theta )e^{i\left( kt+m\varphi \right) },  \notag
\end{eqnarray}%
where $m$ is the azimuthal quantum number and $k$ is the frequency of the
Dirac fields, which is assumed to be positive and real .Since $\left\{
f_{1},f_{2},g_{1},g_{2}\right\} $ and $\left\{
Y_{1},Y_{2},Y_{3},Y_{4}\right\} $ are functions of $r$ and $\theta ,$
respectively, by substituting Eq.(61) into Eq.(60) and applying the
assumptions given by%
\begin{eqnarray}
\text{\ }f_{1}(r) &=&g_{2}(r)\text{ \ \ \ \ and \ \ \ }f_{2}(r)=g_{1}(r)%
\text{\ \ }, \\
Y_{1}(\theta ) &=&Y_{3}(\theta )\text{ \ \ \ \ and \ \ \ }Y_{2}(\theta
)=Y_{4}(\theta ),
\end{eqnarray}%
the Dirac equations transform into Eq.(64). In order to solve the radial
equations , the separation constant $\lambda $ should be defined. This is
achieved by using the angular equations. In fact, it is already known from
the literature that the separation constant can be expressed in terms of the
spin-weighted spheroidal harmonics. The radial parts of the Dirac equations
become

\begin{gather}
\left( \mathbf{D}_{0}+\frac{1}{r}\right) f_{1}\left( r\right) =\frac{\lambda 
}{r\sqrt{2}}f_{2}\left( r\right) , \\
\frac{B\left( r\right) }{2}\left( \mathbf{D}_{0}^{\dagger }+\frac{%
B^{^{\prime }}\left( r\right) }{2B\left( r\right) }+\frac{1}{r}\right)
f_{2}\left( r\right) =\frac{\lambda }{r\sqrt{2}}f_{1}\left( r\right) . 
\notag
\end{gather}%
We further assume that

\begin{eqnarray*}
f_{1}\left( r\right) &=&\frac{\Psi _{1}\left( r\right) }{r}, \\
f_{2}\left( r\right) &=&\frac{\Psi _{2}\left( r\right) }{r},
\end{eqnarray*}%
then Eq.(64) transforms into,

\begin{gather}
\mathbf{D}_{0}\Psi _{1}=\frac{\lambda }{r\sqrt{2}}\Psi _{2}, \\
\frac{B\left( r\right) }{2}\left( \mathbf{D}_{0}^{\dagger }+\frac{%
B^{^{\prime }}\left( r\right) }{2B\left( r\right) }\right) \Psi _{2}=\frac{%
\lambda }{r\sqrt{2}}\Psi _{1}.  \notag
\end{gather}%
Note that $\sqrt{\frac{B\left( r\right) }{2}}\mathbf{D}_{0}^{\dagger }\sqrt{%
\frac{B\left( r\right) }{2}}=\mathbf{D}_{0}^{\dagger }+\frac{B^{^{\prime
}}\left( r\right) }{2B\left( r\right) }+\frac{1}{r}$, and using this
together with the new functions%
\begin{eqnarray*}
R_{1}\left( r\right)  &=&\Psi _{1}\left( r\right) , \\
R_{2}\left( r\right)  &=&\sqrt{\frac{B\left( r\right) }{2}}\Psi _{2}\left(
r\right) ,
\end{eqnarray*}%
and defining the tortoise coordinate $r_{\ast }$ as%
\begin{equation}
\frac{d}{dr_{\ast }}=B\frac{d}{dr},
\end{equation}
Eqs.(65) become%
\begin{eqnarray}
\left( \frac{d}{dr_{\ast }}+ik\right) R_{1} &=&\frac{\sqrt{B}\lambda }{r}%
R_{2}, \\
\left( \frac{d}{dr_{\ast }}-ik\right) R_{2} &=&\frac{\sqrt{B}\lambda }{r}%
R_{1},  \notag
\end{eqnarray}%
In order to write  Eq.(67) in a more compact form, we combine the solutions
in the following way:%
\begin{eqnarray*}
Z_{+} &=&R_{1}+R_{2}, \\
Z_{-} &=&R_{2}-R_{1}.
\end{eqnarray*}%
After doing some calculations we end up with a pair of one-dimensional Schr%
\"{o}dinger-like wave equations with effective potentials,

\begin{gather}
\left( \frac{d^{2}}{dr_{\ast }^{2}}+k^{2}\right) Z_{\pm }=V_{\pm }Z_{\pm },
\\
V_{\pm }=\left[ \frac{B\lambda ^{2}}{r^{2}}\pm \lambda \frac{d}{dr_{\ast }}%
\left( \frac{\sqrt{B}}{r}\right) \right] .
\end{gather}%
In analogy with equation (16), the radial operator $A$ for the Dirac
equations can be written as,

\begin{equation*}
A=-\frac{d^{2}}{dr_{\ast }^{2}}+V_{\pm },
\end{equation*}%
If we write the above operator in terms of the usual coordinates $r$ by
using Eq.(66), we have

\begin{equation}
A=-\frac{d^{2}}{dr^{2}}-\frac{B^{^{\prime }}}{B}\frac{d}{dr}+\frac{1}{B^{2}}%
\left[ \frac{B\lambda ^{2}}{r^{2}}\pm \lambda B\frac{d}{dr}\left( \frac{%
\sqrt{B}}{r}\right) \right] ,
\end{equation}%
Our aim now is to show whether this radial part of the Dirac operator is
essentially self-adjoint or not. This will be achieved by considering 
Eq.(20) and counting the number of solutions that do not belong to Hilbert
space. Hence, Eq.(20) becomes

\begin{equation}
\left( \frac{d^{2}}{dr^{2}}+\frac{B^{^{\prime }}}{B}\frac{d}{dr}-\frac{1}{%
B^{2}}\left[ \frac{B\lambda ^{2}}{r^{2}}\pm \lambda B\frac{d}{dr}\left( 
\frac{\sqrt{B}}{r}\right) \right] \mp i\right) \psi (r)=0.
\end{equation}%
For the asymptotic case, $r\rightarrow \infty $ , the above equation
transforms to 
\begin{equation}
\frac{d^{2}\psi }{dr^{2}}+\frac{2}{r}\frac{d\psi }{dr}=0,
\end{equation}%
whose solution is%
\begin{equation}
\psi \left( r\right) =C_{1}+\frac{C_{2}}{r}.
\end{equation}%
Clearly the solution is square integrable if $C_{1}=0$. Hence, the solution
is asmptotically well behaved. Near $r\rightarrow 0$ , Eq.(71) becomes

\begin{gather}
\frac{d^{2}\psi }{dr^{2}}-\frac{2}{r}\frac{d\psi }{dr}+\frac{\sigma }{r^{3}}%
\psi =0, \\
\sigma =\mp 2\lambda q,  \notag
\end{gather}%
whose solution is given by%
\begin{equation}
\psi \left( r\right) =\left( \frac{4\sigma }{x^{2}}\right) ^{\frac{3}{2}%
}\left\{ C_{3}J_{3}\left( x\right) +C_{4}N_{3}\left( x\right) \right\} ,
\end{equation}%
where $J_{3}\left( x\right) $ and $N_{3}\left( x\right) $ are Bessel
functions of the first and second kind,  and $x=2\sqrt{\frac{\sigma }{r}}.$
As $r\rightarrow 0,$ we have  $x\rightarrow \infty .$ The behavior of the
Bessel functions for real $\nu \geq 0$ as $x\rightarrow \infty $ is given by%
\begin{eqnarray}
J_{\nu }\left( x\right)  &\simeq &\sqrt{\frac{2}{\pi x}}\cos \left( x-\frac{%
\nu \pi }{2}-\frac{\pi }{4}\right) , \\
N_{\nu }\left( x\right)  &\simeq &\sqrt{\frac{2}{\pi x}}\sin \left( x-\frac{%
\nu \pi }{2}-\frac{\pi }{4}\right) ;  \notag
\end{eqnarray}%
thus the Bessel functions asymptotically behave as $J_{3}\left( x\right)
\sim \sqrt{\frac{2}{\pi x}}\cos \left( x-\frac{7\pi }{4}\right) $ and $%
N_{3}\left( x\right) \sim \sqrt{\frac{2}{\pi x}}\sin \left( x-\frac{7\pi }{4}%
\right) .$ Checking for the square integrability has revealed that both
solutions are square integrable. Hence, the radial operator of the Dirac
field fails to satisfy a unique self-adjoint extension condition. As a
result, the occurrence of the timelike naked singularity in the context of $%
f(R)$ gravity remains singular from the quantum mechanical point of view if
it is probed with fermions.

\section{Conclusion}

In this paper, the formation of the naked singularity in the context of a
model of  $f(R)$ gravity is investigated within the framework of quantum
mechanics, by probing the singularity with the quantum fields obeying the
Klein$-$Gordon, Maxwell and Dirac equations. We have investigated the
essential self-adjointness of the spatial part of the wave operator $A$ in
the natural Hilbert space of quantum mechanics which is a linear function
space with square integrability. Our analysis has shown that the timelike
naked curvature singularity remains quantum mechanically singular against
the propagation of the aforementioned quantum fields. Another notable
outcome of our analysis is that the spin of the fields is not effective in
healing of the naked singularity for the considered model of the $f(R)$
gravity spacetime.

Another alternative function space for analyzing the singularity in this
context is to use the Sobelov space instead of the natural Hilbert space 
\cite{13}. The Analysis in Sobelov space entails square integrability both
of the wave function and its derivative. Although the details are not given
in this study, the analysis using the Sobelov space has revealed that
irrespective of the spin structure of the fields used to probe the
singularity, the model considered of $f(R)$ gravity spacetime remains
quantum mechanically singular.

Hence, the generic conclusion that has emerged from our analysis is that in
the model considered of  $f(R)$ gravity, the formation of a timelike naked
singularity is quantum mechanically singular.

It will be interesting for future research to extend the quantum singularity
analysis in other ETG models. Furthermore, it will be a great achievement if
the criterion proposed by HM is extended to stationary metrics. Although the
preliminary work in this direction is considered in \cite{31}, the
formulation has not been fully completed.


\begin{thebibliography}{99}
\bibitem{1} S. Nojiri and S.D. Odintsov, Phys. Rep. \textbf{505}, 59 (2011).

\bibitem{2} T. Clifton and J. D. Barrow, Phys. Rev. D \textbf{72}, 103005
(2005).

\bibitem{3}  T. Multamaki and I. Vilja, Phys. Rev. D \textbf{76}, 064021
(2007). 

\bibitem{4} M. D. Seifert, Phys. Rev. D \textbf{76}, 064002 (2007).

\bibitem{5}  L. Hollenstein and F. S. N. Lobo, Phys. Rev. D \textbf{78},
124007 (2008).

\bibitem{6} A. de la Cruz-Dombriz, A. Dobado and A. L. Maroto, Phys. Rev. D 
\textbf{80}, 124011 (2009).

\bibitem{7} S. Habib Mazharimousavi, M. Halilsoy and T. Tahamtan, Eur. Phys.
J. C, \textbf{72}, 1851 (2012).

\bibitem{8} M. Barriola and A. Vilenkin, Phys. Rev. Lett. \textbf{63}, 341
(1989).

\bibitem{9} P. S. Letelier, Phys. Rev. D \textbf{20}, 1294 (1979).

\bibitem{10} S. Capozziello and M. De Laurentis, Phys. Rep. \textbf{509},
167 (2011).

\bibitem{11} R. M. Wald, J. Math. Phys. (N.Y.) \textbf{21}, 2082 (1980).

\bibitem{12} G. T. Horowitz and D. Marolf, Phys. Rev. D \textbf{52}, 5670
(1995).

\bibitem{13} A. Ishibashi and A. Hosoya, Phys. Rev. D \textbf{60}, 104028
(1999).

\bibitem{14} D. A. Konkowski and T. M. Helliwell, Gen. Rel. and Grav. 
\textbf{33}, 1131, (2001).

\bibitem{15} T. M. Helliwell, D. A. Konkowski and V. Arndt, Gen. Rel. and
Grav. \textbf{35}, 79, (2003).

\bibitem{16} D. A. Konkowski, T. M. Helliwell and C. Wieland, Class. Quantum
Grav. \textbf{21,} 265 (2004).

\bibitem{17} D. A. Konkowski, C. Reese, T. M. Helliwell and C. Wieland, "
Classical and Quantum Singularities of Levi-Civita Spacetimes with and
without a Cosmological Constant", in Procedings of the Workshop on the
Dynamics and Thermodynamics of Black holes and Naked Singularities, ed.
L.Fatibene, M. Francaviglia, R. Giambo and G. Megli, 2004.

\bibitem{18} D. A. Konkowski and T. M. Helliwell, International Journal of
Modern Physics A, Vol.\textbf{26}, No.22, 3878-3888 (2011).

\bibitem{19} J. P. M. Pitelli and P. S. Letelier, J. Math. Phys. \textbf{\
48,} 092501, (2007).

\bibitem{20} J. P. M. Pitelli and P. S. Letelier, Phys. Rev. D \textbf{77,}
124030 (2008).

\bibitem{21} J. P. M. Pitelli and P. S. Letelier, Phys. Rev. D \textbf{80,}
104035 (2009).

\bibitem{22} P. S. Letelier and J. P. M. Pitelli, Phys. Rev. D \textbf{82,}
104046 (2010).

\bibitem{23} O. Unver and O. Gurtug, Phys. Rev. D \textbf{82,} 084016 (2010).

\bibitem{24} S. Habib Mazharimousavi, O. Gurtug and M. Halilsoy, Int. J.
Mod. Phys. D\textbf{18, } 2061-2082, (2009).

\bibitem{25} S. Habib Mazharimousavi, M. Halilsoy, I. Sakalli and O. Gurtug,
Class. Quant. Grav. \textbf{27}, 105005, (2010).

\bibitem{26} S. Habib Mazharimousavi, O. Gurtug, M. Halilsoy and O. Unver,
Phys. Rev. D \textbf{84,} 124021 (2011).

\bibitem{27} M. Reed and B. Simon, \textit{Functional Analysis}, (Academic
Press, New York, 1980).

\bibitem{28} M. Reed and B. Simon, \textit{Fourier Analysis and
Self-Adjointness}, (Academic Press, New York, 1975).

\bibitem{29} R. D. Richtmyer, \textit{Principles of Advanced Mathematical
Physics}, (Springer, New York, 1978).

\bibitem{30} S. Chandrasekhar, \textit{The Mathematical Theory of Black Holes%
}, (Oxford University Press, 1992).

\bibitem{31} I. Seggev, Class. Quant. Grav. \textbf{21}, 2651, (2004).
\end{thebibliography}
\end{document}